\documentclass[11pt,twoside]{article}
\usepackage{amssymb}

\def\ii{\'{\i}}                     

\def\r{\mathbb R}                   
\def\n{\mathbb N}                   
\def\L{\mathbb L}                   

\def\rw{\mathbb R\mathbb W}

\newcommand{\bm}[1]{\mbox{\boldmath $#1$}}
\def\be{\begin{equation}}
\def\ee{\end{equation}}
\def\bea{\begin{eqnarray}}
\def\eea{\end{eqnarray}}
\def\b*{\begin{eqnarray*}}
\def\e*{\end{eqnarray*}}
\def\N{\hfill \rule{2.5mm}{2.5mm}}

\def\g{\gamma}
\def\z{\zeta}
\def\f{\varphi}
\def\u{\vec u}
\def\k{\vec k}
\def\x{\vec x}
\def\y{\vec y}
\def\C{{\cal C}}

\def\Z{\Theta}
\def\DP{{\cal DP}}

\def\P{{\it Proof:} \hspace{3mm}}

\def\xiv{\vec \xi }
\def\lie{{\pounds}_{\xiv}\, }

\newtheorem{defi}{Definition}[section]
\newtheorem{theo}{Theorem}[section]
\newtheorem{coro}{Corollary}[section]
\newtheorem{prop}{Proposition}[section]
\newtheorem{lem}{Lemma}[section]
\newtheorem{pr}{Property}[section]
\newtheorem{rem}{Remark}[section]

\begin{document}

\pagestyle{myheadings}
\markboth
{{\sc Superenergy tensors }}  
{{\sc J M M Senovilla  }}        

\vspace{1cm}




\thispagestyle{empty}

\begin{center}

{\Large{\bf{Superenergy tensors and their applications}}} 



\vskip1cm

{\sc Jos\'e M. M. Senovilla}   

\vskip0.5cm

{\it Departamento de F\ii sica Te\'orica, Universidad del Pa\ii s 
Vasco,\\
Apartado 644, 48080 Bilbao, Spain} 

\end{center}
\bigskip

%
%

\begin{abstract}
In Lorentzian manifolds of any dimension the concept of
{\em causal tensors} is introduced. Causal tensors have positivity properties
analogous to 
the so-called ``dominant energy condition''. Further, it is shown how 
to build, from \underline{any} given tensor $A$, a new tensor quadratic 
in $A$ and ``positive'', in the sense that it is causal. These tensors 
are called {\em superenergy tensors} due to historical reasons because 
they generalize the classical energy-momentum and Bel-Robinson 
constructions. Superenergy tensors are basically unique and with 
multiple and diverse physical and mathematical applications, such as: 
a) definition of new divergence-free currents, b) conservation laws 
in propagation of discontinuities of fields, c) the causal propagation 
of fields, d) null-cone preserving maps, e) generalized Rainich-like 
conditions, f) causal relations and transformations, and g) generalized 
symmetries. Among many others.
\end{abstract}


\section{Causal tensors}
In this contribution\footnote{It is worthwhile to check also my joint
contribution with Garc\ii a-Parrado, as well as that of Bergqvist's,
in this volume, with related results. Notice that signature convention here
is opposite to those contributions and to \cite{BS}.}
$V$ will denote a differentiable 
$N$-dimensional manifold $V$ endowed with a metric $g$ of Lorentzian signature
$N-2$. The solid Lorentzian cone at $x$ will be denoted by $\Z_{x}=\Z^+_{x}\cup 
\Z^-_{x}$, where $\Z^{\pm}_{x}\subset T_{x}(V)$ are the future (+) and 
past (--) half-cones. The null cone $\partial \Z_{x}$ is the boundary of $\Z_{x}$ 
and its elements are the null vectors at $x$. 
An arbitrary point $x\in V$ is usually taken, but all 
definitions and results translate immediately to tensor fields if a 
time orientation has been chosen. The $x$-subscript is then dropped.
\begin{defi}{\em \cite{BS}}
A rank-r tensor $T$ has the \underline{dominant property} at $x\in V$ if
$$T(\vec{u}_{1},\dots ,\vec{u}_{r})\geq 0
\hspace{1cm} \forall \vec{u}_{1},\dots ,\u_{r}\in \Z^+_{x}.$$
The set of rank-$r$ tensor (fields) with the dominant property will be denoted by 
$\DP^+_{r}$. We also put
$\DP^-_{r}\equiv \{T:\, -T\in \DP^+_{r} \}, \hspace{3mm} \DP_{r}\equiv 
\DP^+_{r}\cup\DP^-_{r}, \hspace{3mm} 
\DP^{\pm}\equiv\bigcup_{r}\DP^{\pm}_{r}, \hspace{3mm} \DP\equiv 
\DP^+\cup\DP^-.$
\label{DP}
\end{defi}
By a natural extension $\r^+ =\DP_{0}^{+}\subset \DP^+$.
Rank-1 tensors in $\DP^+$ are simply the past-pointing
causal 1-forms (while those in $\DP^-_{1}$ are the future-directed ones). For
rank-2 tensors, the dominant property was introduced by Pleba\'{n}ski
\cite{Ple} in General Relativity and is usually called the dominant energy
condition \cite{HE} because it is a requirement for physically
acceptable energy-momentum tensors. The elements of $\DP$ will be called
``\underline{causal tensors}''.
As in the case of past- and future-pointing vectors, any statement
concerning $\DP^+$ has its counterpart concerning $\DP^-$, and they will be
generally omitted. Trivially one has
\begin{pr}
If $T^{(i)}\in \DP^+_{r}$ and $\alpha _{i}\in \r^+$ ($i=1,...,n$)
then $\sum\limits_{i=1}^{n}\alpha _{i}T^{(i)}\in\DP^+_{r}$.
Moreover, if
$T^{(1)}$ , $T^{(2)}\in \DP^+$ then
$T^{(1)}\otimes T^{(2)} \in \DP^+$.
\label{pr:alg}
\end{pr}
This tells us that $\DP^+$ is a graded algebra of cones. 
For later use, let us introduce the following notation
$$
T^{(1)}\, {}_i\!\times_j\, T^{(2)}\equiv
C^1_{i}C^2_{r+j} \left(g^{-1}\otimes T^{(1)}\otimes T^{(2)}\right)
$$
that is to say, the contraction (via the metric) of the $i^{th}$ entry of
the first tensor (which has rank $r$) with the $j^{th}$ of the second.
There are of course many different products
${}_i\!\times_j$ depending on where the contraction is made.

Several characterizations of $\DP^+$ can be found. For instance 
\cite{BS,S}
\begin{prop}
The following conditions are equivalent:
\begin{enumerate}
    \item $T\in \DP^+_{r}$.
    \item $T(\k_{1},\dots ,\k_{r})\geq 0
\hspace{3mm} \forall \k_{1},\dots ,\k_{r}\in \partial \Z^+_{x}$.
    \item $T(\u_{1},\dots ,\u_{r})> 0
\hspace{3mm} \forall \u_{1},\dots ,\u_{r}\in \mbox{\rm{int}} 
\Z^+_{x}$, ($T\neq 0$).
\item $T(\vec{e}_{0},\dots,\vec{e}_{0})\geq \left|
T(\vec{e}_{\alpha _{1}},...,\vec{e}_{\alpha_{r}})\right| 
\hspace{1mm} \forall \alpha_{1},...,\alpha_{r}\in \left\{0,1,...,N-1\right\}$,
in {\em all} orthonormal bases $\left\{\vec{e}_{0},...,\vec{e}_{N-1}\right\}$
with a future-pointing timelike $\vec{e}_{0}$.
\item $T\, {}_i\!\times_j\, \tilde{T} \in \DP^+_{r+s-2},
\hspace{3mm} \forall \tilde{T}\in \DP^-_{s},\,\, \forall i=1,\dots ,r, \, \,
\forall j=1,\dots ,s.$
\end{enumerate}
\label{dp}
\end{prop}
\begin{prop}
    Similarly, some characterizations of $\DP$ are \cite{BS}
\begin{enumerate}
    \item $0\neq T{}_i\!\times_i\! T\in \DP^-$ for some
$i \Longrightarrow T{}_i\!\times_j\! T \in \DP^-$ for {\em all} $i,j
\Longrightarrow T\in\DP$.
\item $T{}_i\!\times_i\! T=0$ for {\em all}
$i \Longleftrightarrow T=k_{1}\otimes\dots\otimes k_{r}$ where $k_{i}$
are null $\Longrightarrow  T\in \DP$.
\end{enumerate}
\end{prop}


\section{Superenergy tensors}
In this section the questions of how general is the class $\DP$ and how
one can build causal tensors are faced. The main result is that:

{\em Given an {\em arbitrary} tensor $A$, there is a canonical procedure
(unique up to permutations) to construct a \underline{causal} tensor
quadratic in $A$.}

This procedure was introduced in \cite{S1} and extensively considered in
\cite{S}, and the causal tensors thus built are called
``super-energy tensors''. The whole thing is based in the following 
\begin{rem}
Given any rank-$m$ tensor $A$, there is a {\em minimum} value 
$r\in \n$, $r\leq m$ and a unique set of $r$ numbers $n_{1},\dots 
,n_{r}\in\n$, with $\sum_{i=1}^{r}n_{i}=m$, such that $A$ is a linear 
map on $\Lambda^{n_{1}}\times\dots\times \Lambda^{n_{r}}$.
\end{rem}
Here $\Lambda^{p}$ stands for the vector space of ``contravariant $p$-forms''
at any $x\in V$. In other words, $\exists$ a minimum $r$ such that
$\tilde A\in \Lambda_{n_{1}}\otimes\dots\otimes \Lambda_{n_{r}}$, 
where $\tilde A$ is the appropriate permutted version of $A$ which 
selects the natural order for the $n_{1},\dots ,n_{r}$ entries. Tensors 
seen in this way are called $r$-fold $(n_{1},\dots ,n_{r})$-forms. 
Some simple examples are: any $p$-form $\Omega$ is trivially a 
single (that is, 1-fold) $p$-form, while $\nabla\Omega$ is a double 
$(1,p)$-form; the Riemann tensor $R$ is a double (2,2)-form which is 
symmetric (the pairs can be interchanged), while $\nabla R$ is a 
triple (1,2,2)-form; the Ricci tensor $Ric$ is a double symmetric 
(1,1)-form and, in general, any completely symmetric $r$-tensor is 
an $r$-fold (1,1,\dots ,1)-form. A 3-tensor $A$ with the property 
$A(\vec x, \vec y, \vec z)=-A(\vec z, \vec y, \vec x)$ is a double 
(2,1)-form and the corresponding $\tilde A$ is clearly given by 
$\tilde A (\vec x, \vec y, \vec z)\equiv A(\vec x, \vec z, \vec y), 
\hspace{1mm} \forall \vec x, \vec y, \vec z$.

For $r$-fold forms, the interior contraction can be generalized in the 
obvious way $i_{\x_{1},\dots ,\x_{r}}:
\Lambda_{n_{1}}\otimes\dots\otimes \Lambda_{n_{r}} \longrightarrow
\Lambda_{n_{1}-1}\otimes\dots\otimes \Lambda_{n_{r}-1}$ by means of
$$
i_{\x_{1},\dots ,\x_{r}}A=C^1_{1}C^2_{n_{1}+1}\dots 
C^r_{n_{1}+\dots +n_{r-1}+1}\, \left(\x_{1}\otimes 
\x_{2}\otimes\dots\otimes\x_{r}\otimes\tilde A\right)
$$
which is simply the interior contraction of each vector with each 
antysymmetric block. Similarly, by using the canonical volume element
of $(V,g)$ one can define the multiple Hodge duals as follows:
$$
*_{\cal P}:
\Lambda_{n_{1}}\otimes\dots\otimes \Lambda_{n_{r}} \longrightarrow
\Lambda_{n_{1}+\epsilon_{1}(N-2n_{1})}\otimes\dots\otimes
\Lambda_{n_{r}+\epsilon_{r}(N-2n_{r})} 
$$
where $\epsilon_{i}\in \{0,1\}\, \forall i=1,\dots ,r$ and the convention
is taken that $\epsilon_{i}=1$ if the $i^{th}$ antysymmetric 
block is dualized and $\epsilon_{i}=0$ otherwise, and
where ${\cal P}=1,\dots ,2^r$ is defined by
${\cal P}=1+\sum_{i=1}^r 2^{i-1}\epsilon_{i}$. Thus, there are $2^r$ 
different Hodge duals for any $r$-fold form $A$ and they can be 
adequately written as $A_{\cal P}\equiv *_{\cal P}A$. One also needs
a product $\odot$ of $A$ by itself
resulting in a $2r$-covariant tensor, given by
$$
(A\odot A)\left(\x_{1},\y_{1},\dots ,\x_{r},\y_{r}\right)\equiv 
\left(\prod_{i=1}^{r}\frac{1}{(n_{i}-1)!}\right)
g\left(i_{\x_{1},\dots ,\x_{r}}A,i_{\y_{1},\dots ,\y_{r}}A\right)
$$
where for any tensor $B$ we write $g\left(B,B\right)\in \r$ for the complete 
contraction in all indices in order.
\begin{defi}{\em \cite{S,S1}}
The basic \underline{superenergy tensor} of $A$ is defined to be
$$
T \{ A\} ={1\over 2}\sum_{{\cal P}=1}^{2^r}
A_{\cal P}\odot A_{\cal P} .
$$
\label{def:se}
\end{defi}
Here the word basic is used because linear combinations of $T\{A\}$ 
with its permutted versions maintain most of its properties; however, 
the completely symmetric part is unique (up to a factor of proportionality)
\cite{S}. It is remarkable that one can provide an explicit expression for
$T\{A\}$ which is {\it independent} of the dimension $N$, see \cite{S}.
In the case of a general $p$-form $\Omega$, its rank-2
superenergy tensor becomes 
\be
T\{\Omega\}\left(\x ,\y\right)=\frac{1}{(p-1)!}\left[
g\left(i_{\x}\Omega ,i_{\y}\Omega\right)-
\frac{1}{2p}g\left(\Omega , \Omega\right) g\left(\x ,\y\right)
\right].\label{sep}
\ee
In Definition \ref{def:se} we implicitly assumed that the $r$-fold 
form $A$ has no antysymmetric blocks of maximum degree $N$. 
Nevertheless, the above expression (\ref{sep}) is perfectly well defined
for $N$-forms: if $\Omega=f\eta$ where
$\eta$ is the canonical volume form and $f$ a scalar, then (\ref{sep}) gives
$T\{\Omega \}=-{1\over 2}f^2 g$. Using this the Definition
\ref{def:se} is naturally extended to include $N$-blocks, see 
\cite{BS} for details.

In $N=4$, the superenergy tensor of a 2-form $F$ is its
Maxwell energy-momentum tensor, and the superenergy tensor of an exact
1-form $d\phi$ has the form of the energy-momentum tensor for a massless
scalar field $\phi$. If we compute the superenergy tensor of $R$ we get
the so-called Bel tensor \cite{Bel}.
The superenergy tensor of the Weyl curvature tensor
is the well-known Bel-Robinson tensor \cite{B,Bel2,MTW}. 
The main properties of $T\{A\}$ are \cite{S}:
\begin{pr}
If $A$ is an $r$-fold form, then $T\{A\}$ is a $2r$-covariant tensor.
\end{pr}
\begin{pr}
$T\{A\}$ is symmetric on each pair of entries, that is, for all 
$i=1,\dots,r$ one has
$$
T\{A\}\left(\x_{1},\dots,\x_{2i-1},\x_{2i},\dots,\x_{2r}\right)=
T\{A\}\left(\x_{1},\dots,\x_{2i},\x_{2i-1},\dots,\x_{2r}\right).
$$
\end{pr}
\begin{pr}
$T\{A\}=T\{A_{\cal P}\} \,\,\, \forall {\cal P}=1,\dots ,2^r$.
\end{pr}
\begin{pr}
$T\{A\}=T\{-A\}$; \,\, 
$T\{A\}=0 \Longleftrightarrow A=0$.\label{cero}
\end{pr}
\begin{pr}
$T\{A\otimes B\}=T\{A\}\otimes T\{B\}$.\label{tensorprod}
\end{pr}
\begin{pr}
$T\{A\}\in \DP^+$.\label{seindp}
\end{pr}
Observe that property \ref{seindp} is what we were seeking, so that 
$T\{A\}$ is the ``positive square'' of $A$ in the causal sense.
\begin{pr}
$\displaystyle{T\{A\}(\vec{e}_{0},\dots,\vec{e}_{0})=\frac{1}{2} 
\sum_{\alpha_{1},...,\alpha_{m}=0}^{N-1}
\left(A(\vec{e}_{\alpha _{1}},...,\vec{e}_{\alpha_{m}})\right)^2}$
in orthonormal bases $\left\{\vec{e}_{0},...,\vec{e}_{N-1}\right\}$.
\end{pr}

The set of superenergy tensors somehow build up the class $\DP$; in 
fact, in many occasions the rank-2 superenergy tensors (that is, those 
for single $p$-forms) are the basic building blocks of the whole 
$\DP$ \cite{BS}. This can be seen as follows.
\begin{defi}
An $r$-fold form $A$ is said to be \underline{decomposable}
if there are $r$ forms $\Omega_{i}$ ($i=1,\dots ,r$)
such that $\tilde{A}=\Omega_{1}\otimes \dots \otimes \Omega_{r}$.
\label{def:decom}
\end{defi}
 From this and Property \ref{tensorprod} one derives
\begin{coro}
If $A$ is decomposable, then $T\left\{A\right\}=
T\left\{\Omega_{1} \right\}\otimes \dots \otimes
T\left\{\Omega_{r}\right\}.$
\label{cor:decom}
\end{coro}
Notice that each of the $T\left\{\Omega_{i} \right\}$ on the 
righthand side is a rank-2 tensor. We now have
\begin{theo}
Any symmetric $T\in \DP^+_2$
can be decomposed as
$$T=\sum_{p=1}^N T\{\Omega_{p}\}$$
where $\Omega_{p}$ are \underline{simple} $p$-forms
such that for $p>1$ they have
the structure $\Omega_{p}=k_{1}\wedge \dots \wedge k_{p}$
where $k_{1},\dots ,k_{p}$ are appropriate
{\em null} 1-forms and $\Omega_{1}\in \DP_{1}$.
\label{theo:bb2}
\end{theo}
See \cite{BS} for the detailed structure of the above decomposition and 
for the relation between $T$ and the null 1-forms. From this one 
obtains\footnote{From this point on I shall use the standard 
notation $T^2$ instead of $T{}_1\!\times_1\! T=T{}_2\!\times_2\! T
=T{}_2\!\times_1\! T=T{}_1\!\times_2\! T$
for the case of rank-2 symmetric tensors $T$. $T^2$ is symmetric.}
\begin{theo}
A symmetric rank-2 tensor $T$ satisfies $T^2 =g$ if 
and only if $T=\pm T\{\Omega_{p}\}$, i.e., if $T$ is up to sign
the superenergy tensor of a simple
$p$-form $\Omega_{p}$. Moreover, the rank $p$ of the $p$-form is given 
by $\pm${\rm tr}$T=2p-N$.\label{txt}
\end{theo}
This important theorem allows to classify all Lorentz transformations 
and, in more generality, all maps which preserve the null cone
$\partial \Z_{x}$ at $x\in V$, see \cite{BS}.


\section{Applications}
In this section several applications of superenergy and causal tensors 
are presented. They include both mathematical and physical ones.
\subsection{Rainich's conditions}
The classical Rainich conditions \cite{R,MW} are necessary and sufficient 
conditions for a spacetime to originate via Einstein's equations in a 
Maxwell electromagnetic field. They are of two kinds: algebraic and 
differential. Here I am only concerned with the algebraic part which 
nowadays are presented as follows (see, e.g., \cite{PR}):

(Classical Rainich's conditions) {\em The Einstein tensor $G=Ric 
-\frac{1}{2}S\, g$ of a 4-dimensional spacetime is proportional to the 
energy-momentum tensor of a Maxwell field (a 2-form) if and only if
$G^2\propto g$, {\rm tr}$G=0$ and $G\in DP^+_{2}$.}

In fact, from theorem \ref{txt} one can immediately
improve a little this classical result
\begin{coro}
In $4$ dimensions, $G$ is {\em algebraically} up to sign proportional 
to the energy-momentum tensor of a Maxwell field 
if and only if $G^2\propto g$, {\rm tr}$G=0$. Furthermore, $G$ 
is proportional to the energy-momentum tensor of (possibly another) 
Maxwell 2-form which is {\em simple}.
\end{coro}
The last part of this corollary is related to the so-called duality 
rotations of the electromagnetic field \cite{PR}. Observe that this is 
clearly a way of determining physics from geometry because, given a 
particular spacetime, one only has to compute its Einstein tensor and 
check the above simple conditions. If they hold, then the 
energy-momentum tensor is that of a 2-form (and for a complete result 
the Rainich differential conditions will then be needed).

The classical Rainich conditions are based on a dimensionally-dependent
identity, see \cite{L}, valid only for $N=4$. However, theorem 
\ref{txt} has universal validity and can be applied to obtain the 
generalization of Rainich's conditions in many cases. For 
instance, we were able to derive the following results \cite{BS}.
\begin{coro}
In $N$ dimensions, a rank-2 symmetric tensor $T$ is {\em algebraically}
the energy-momentum tensor of a minimally coupled massless scalar field
$\phi$ if and only if $T^2\propto g$ and {\rm tr}$T =\beta
\sqrt{{\rm tr}T^2/N}$
where $\beta =\pm (N-2)$. Moreover, $d\phi$ is spacelike if $\beta =2-N$
and {\rm tr}$T\ne 0$, timelike if $\beta =N-2$ and {\rm tr}$T\ne 0$, and null
if {\rm tr}$T =0=${\rm tr}$T^2$.
\end{coro}
\begin{coro}
In $N$ dimensions, a rank-2 symmetric tensor $T$ is
the energy-momentum tensor of a perfect fluid satisfying
the dominant energy condition if and only if there exist two positive 
functions $\lambda,\mu$ such that 
$$
T^2=2\lambda T+\left(\mu^2-\lambda^2\right) g,
\hspace{15mm} \mbox{{\rm tr}}T= (N-2)\mu\, -\lambda\, N.
$$
\end{coro}
This is an improvement and a generalization to arbitrary $N$
of the conditions in \cite{CF} for $N=4$.
In particular, the case of dust can be deduced from the previous one
by setting the pressure of the perfect fluid equal to zero.
\begin{coro}
In $N$ dimensions, a symmetric tensor $T$ is {\em algebraically}
the energy-momentum tensor of a dust satisfying the dominant
energy condition if and only if  
$$
T^2= (\mbox{{\rm tr}}T)\, T,
\hspace{15mm} \mbox{{\rm tr}}T<0 .
$$
\end{coro}

\subsection{Causal propagation of fields}
Following a classical reasoning appearing in \cite{HE}, the causal 
propagation of arbitrary fields can be studied by simply using its 
superenergy tensor, see \cite{BerS}. Let $\z$ be any closed achronal set in
$V$ and $D(\z)$ its total Cauchy development (an overbar over a set 
denotes its closure, see \cite{HE,S0} for definitions and notation). 
\begin{theo}
If the tensor $T\{A\}$ satisfies the following divergence inequality
$$
\mbox{\rm div}T\left\{A\right\}\left(\x ,\dots ,\x\right)\leq f\,
T\left\{A\right\}\left(\x ,\dots ,\x\right)
$$
where $f$ is a continuous function and $\x=g^{-1}(\hspace{2mm} ,-d\tau)$
is any timelike vector foliating $D(\z)$ with hypersurfaces $\tau=$const.,
then
\b*
\left.A\right|_{\z}=0 \hspace{1cm} \Longrightarrow
\hspace{1cm} \left.A\right|_{\overline{D(\z)}}=0.
\e*
\label{th:causal}
\end{theo}
This theorem proves the causal propagation of the field $A$ because 
if $A\neq 0$ at a point $x\notin \overline{D(\z)}$ arbitrarily close to
$\overline{D(\z)}$, then $A$ will propagate in time from $x$ according to its
field equations, but it will never be able to enter into $\overline{D(\z)}$,
showing that $A$ cannot travel faster than light.

The divergence condition in the theorem, being an inequality, is very 
mild and it is very easy to check whether or not is valid for a given 
field satysfying field equations. In general, of course, it will work 
for linear field equations, and for many other cases too.
It has been used to prove the 
causal propagation of gravity in vacuum \cite{BoS} or in general 
$N$-dimensional Lorentzian manifolds conformally related to Einstein 
spaces \cite{BerS}, and also for the massless spin-$n/2$ fields in 
General Relativity \cite{BerS}. It must be stressed that in many 
occasions the standard energy-momentum tensor of the field does not 
allow to prove the same result, so that the universality of the 
superenergy construction reveals itself as essential in this 
application.

\subsection{Propagation of discontinuities: conserved quantities}
Several ways to derive conserved quantities and exchange of superenergy
properties have been pursued. One of them is the construction of 
divergence-free vector fields, called currents. This has been 
succesfully achieved in the case of a minimally coupled scalar field 
if the Einstein-Klein-Gordon field equations hold, see \cite{S,S2}.
In this subsection the propagation of discontinuities of the 
electromagnetic and gravitational fields will be analyzed from the 
superenergy point of view. This will be enough to prove the interchange
of superenergy quantities between these two physical fields and some
conservation laws arising naturally when the field has a `wave-front', 
see \cite{S,S2}.

To that end, we need to recall some well-known basic properties of the 
wave-fronts, which are null hypersurfaces. Let $\sigma$ be such a null 
hypersurface and $n$ a 1-form normal to $\sigma$. Obviously, $n$ is null
$g^{-1}(n,n)=0$ and therefore ${\vec n}\equiv g^{-1}(\hspace{2mm},n)$ is
in fact a vector {\em tangent} to $\sigma$, see e.g. \cite{MS}, and
$n$ cannot be normalized so that it is defined up to a transformation
of the form
\be
n \hspace{1mm} \longrightarrow \hspace{1mm} \rho \, n\, ,\hspace{15mm} \rho >0. 
\label{free}
\ee
The null curves tangent to $\vec{n}$ are null geodesics
$\nabla_{\vec n}\vec n=\Psi \,  \vec n$, called `bicharacteristics',
contained in $\sigma$.
Let $\bar{g}$ denote the first fundamental form of $\sigma$, which is a 
degenerate metric because $\bar{g}(\vec n, \hspace{2mm})=0$ \cite{HE,MS,S0}.
The second fundamental form of $\sigma$ relative to $n$ can be defined
as
\b*
K\equiv \frac{1}{2} \pounds_{\vec{n}} \, \bar{g}
\e*
where $\pounds_{\vec{n}}$ denotes the Lie derivative with
respect to $\vec{n}$ within $\sigma$. $K$ is {\em intrinsic}
to the null hypersurface $\sigma$ and shares the degeneracy with
$\bar{g}$: $K(\vec n, \hspace{2mm})=0$ \cite{MS}. Because of this, and even
though $\bar{g}$ has no inverse, one can define the ``trace''
of $K$ by contracting with the inverse of the
metric induced on the quotient spaces $T(\sigma)/<\vec n>$.
This trace will be denoted by $\vartheta$ and has the following interpretation:
if $s\subset \sigma$ denotes any spacelike cut of $\sigma$, that is,
a spacelike $(N-2)$-surface orthogonal to $n$ and within $\sigma$, 
then $\vartheta$ measures the volume expansion of $s$ along the
null generators of $\sigma$. In fact, $\vartheta$ can be easily related
to the derivative along $\vec{n}$ of the $(N-2)$-volume element of $s$
\cite{S0}.

Let us consider the case when there is an
electromagnetic field (a 2-form $F$) propagating in a background spacetime
so that there is a {\it null} hypersurface of discontinuity $\sigma$
\cite{Fri,Li}, called a `characteristic'. Let $[E]_{\sigma}$ denote the
discontinuity of any object $E$ across $\sigma$. Using the classical Hadamard
results \cite{Had,Fri,Li}, one can prove the
existence of a 1-form $c$ on $\sigma$ such that
$$
\left[F\right]_{\sigma}=n\wedge c \, , \hspace{3mm} g^{-1}(n,c)=0.
$$
Observe that $c$ transforms under the freedom (\ref{free}) as
$c \hspace{1mm} \longrightarrow \hspace{1mm} c/ \rho$.
From Maxwell's equations for $F$ considered in a distributional sense
one derives a propagation law \cite{Li,S2}, or `transport equation' \cite{Fri},
\b*
\vec n \left(|c|^2\right)+|c^2|(\vartheta +2\Psi)=0,\hspace{5mm}
|c|^2\equiv g^{-1}(c,c)\geq 0.
\e*
This propagation equation implies that if $c|_{x}=0$ at any point $x\in \sigma$,
then $c=0$ along the null geodesic originated at $x$ and tangent
to $\vec{n}$. Moreover, for arbitrary conformal Killing vectors
$\vec{\z}_1,\vec{\z}_2$,
the above propagation law allows to prove that \cite{Fri,S2}
\be
\int_{s} |c|^2 n(\vec{\z}_1)n(\vec{\z}_2)\,
\bm{\omega}|_{s} \label{cq}
\ee
are conserved quantities along $\sigma$, where $\bm{\omega}|_s$ is the canonical
volume element $(N-2)$-form of $s$, in the sense that the integral is 
independent of the cut $s$ chosen. Notice also that (\ref{cq})
are invariant under the transformation (\ref{free}).
These conserved quantities can be easily related to the energy-momentum
properties of the electromagnetic field because
$T\left\{[F]_{\sigma}\right\}=|c|^2 n\otimes n$,
so that the Maxwell tensor of the discontinuity $[F]_{\sigma}$
contracted with the conformal Killing vectors $\vec{\z}_1,\vec{\z}_2$ gives the
function integrated in (\ref{cq}).

However, the integral (\ref{cq}) vanishes when
$\left[F\right]_{\sigma}=0 \Longleftrightarrow c=0$. 
Using again Hadamard's theory, now there exist a 2-covariant
symmetric tensor $B$ and a 1-form $f$ defined only on $\sigma$
such that \cite{Bel2,Ge,Li,MS}
\b*
\left[R\right]_{\sigma}=n\wedge B\wedge n, 
\hspace{5mm} B(\vec n ,\,\,)+\mbox{\rm tr}B\, n=0, \\
\left[\nabla F\right]_{\sigma}= n\otimes (n\wedge f),
\hspace{5mm} g^{-1}(n,f)=0 .
\e*
These objects transform under (\ref{free}) according to $
f,B \longrightarrow f/ \rho^2, B/ \rho^2$.
Assuming that the Einstein-Maxwell equations hold Lichnerowicz
deduced the propagation laws for $f,B$ in \cite{Li}, and in 
particular he found the following transport equation \cite{Li,S}
\b*
\vec n \left(|B|^2+ |f|^2\right)+\left(|B|^2+
|f|^2\right)\left(\vartheta +4\Psi\right)=0,
\e*
where $|f|^2\equiv g^{-1}(f,f)\geq 0, |B|^2\equiv$tr$B^2\geq 0$.
Once again, with the help of any conformal Killing vectors
$\vec{\z}_1,\dots,\vec{\z}_4$, the following quantities 
\be
\int_{s} \left(|B|^2+ |f|^2\right)n(\vec{\z}_1)n(\vec{\z}_2)
n(\vec{\z}_3)n(\vec{\z}_4)\,\, \bm{\omega}|_s \label{em-g}
\ee
are conserved along $\sigma$ in the sense that the integral is
independent of the spacelike cut $s$. Two important points can be 
derived from this relation: first, {\it both} the electromagnetic
and gravitational contributions are necessary,
so that neither the integrals involving only $|B|^2$ or only $|f|^2$
are equal for different cuts $s$ in general.
Second, the integrand in (\ref{em-g}) is related to superenergy
tensors because
$T\left\{[R]_{\sigma}\right\}=
2|B|^2 n\otimes n\otimes n\otimes n$ and
$T\left\{[\nabla F]_{\sigma}\right\}=
2|f|^2 n\otimes n\otimes n\otimes n$, and thus the function integrated
in (\ref{em-g}) is simply
\b*
\frac{1}{2}\left(T\left\{[R]_{\sigma}\right\}+
T\left\{[\nabla F]_{\sigma}\right\}\right)(\vec{\z}_1,\vec{\z}_2,
\vec{\z}_3,\vec{\z}_4)
\e*
which demonstrates the interplay between superenergy quantities
of different fields, in this case the electromagentic and gravitational
ones. Observe that the above tensors are completely symmetric in this
case, and that they together with the conserved quantity
(\ref{em-g}) are invariant under the transformation (\ref{free}).

\subsection{Causal relations}
The fact that the tensors in $\DP_{2}$ can be seen as linear 
mappings preserving the Lorentz cone leads in a natural way to 
consider the possibility of relating different Lorentzian manifolds 
at their corresponding causal levels, even before the metric 
properties are taking into consideration. To that end, using the 
standard notation $\f^{'}$ and $\f^{*}$ for the push-forward and 
pull-back mappings, respectively, we give the next
\begin{defi}
Let $\f:V\rightarrow W$ be a global diffeomorphism between
two Lorentzian manifolds. $W$ is said to be
properly causally related with $V$ by $\f$, denoted 
$V\prec_{\f}W$, if $\forall \,\, \u\in\Z^{+}(V)$, $\f^{'}\u\in \Z^{+}(W)$.
$W$ is simply said to be properly causally related with $V$,
denoted by $V\prec W$, if $\exists \f$ such that $V\prec_{\f}W$.
\label{PREC}
\end{defi}
In simpler terms, what one demands here is that the solid Lorentz 
cones at all $x\in V$ are mapped by $\f$ to sets contained in the 
solid Lorentz cones at $\f (x)\in W$ keeping the time orientation: 
$\f '\Z^{+}_{x}\subseteq \Z^{+}_{\f (x)}, \, \, \, \forall x\in V$.

Observe that two Lorentzian manifolds can be properly causally related 
by some diffeomorphisms but not by others. As a simple example, 
consider $\L$ with typical Cartesisan coordinates $x^{0},\dots 
,x^{N-1}$ (the 0-index indicating the time coordinate) and let 
$\f_{q}$ be the diffeomorphisms
\b*
\f_{q}: \L &\longrightarrow & \L \\
(x^{0},\dots ,x^{N-1}) &\longrightarrow & (q\,x^{0},\dots ,x^{N-1})
\e*
for any constant $q\neq 0$. It is easily checked that $\f_{q}$ is a proper 
causal relation for all $q\geq 1$, but is not for all $q<1$. Thus 
$\L\prec \L$ but, say, $\L\nprec_{\f_{1/2}} \L$. Notice also that for 
$q\leq -1$ the diffeomorphisms $\f_{q}$ change the time 
orientation of the causal vectors, but still
$\f '\Z_{x}\subseteq \Z_{\f (x)}$ (with
$\f '\Z^{+}_{x}\subseteq \Z^{-}_{\f (x)}$.)

Proper causal relations can be easily characterized by some 
equivalent simple conditions.
\begin{theo}
The following statements are equivalent:
\begin{enumerate}
    \item $V\prec_{\f}W$.   
    \item $\f^{*}\left(\DP_{r}^+(W)\right)\subseteq \DP_{r}^+(V)$ for 
    all $r\in \n$.
    \item $\f^{*}\left(\DP_{1}^+(W)\right)\subseteq \DP_{1}^+(V)$.
    \item $\f^{*}h\in \DP_{2}^-(V)$ where $h$ is the metric 
    of $W$ up to time orientation.
\end{enumerate}
\label{pcr}
\end{theo}
\P

\noindent
$1 \Rightarrow 2$: let $T\in \DP_{r}^+(W)$, then 
$(\f^{*}T) (\u_{1},\dots ,\u_{r})=
T\left(\f'\u_{1},\dots ,\f'\u_{r}\right)\geq 0$ for all
$\u_{1},\dots ,\u_{r}\in\Z^+(V)$ 
given that $\f'\u_{1},\dots ,\f'\u_{r}\in \Z^{+}(W)$ by assumption.
Thus $\f^{*}T\in \DP_{r}^+(V)$.

\noindent
$2 \Rightarrow 3$: Trivial.

\noindent
$3 \Rightarrow 4$: Choose any basis $w^1,\dots ,w^{N}$ of $T^{*}(W)$ 
such that $w^1,\dots ,w^{N}\in \DP_{1}^+(W)$. Then
$h=\sum_{\mu ,\nu =1}^{N}h_{\mu\nu}w^{\mu}\otimes w^{\nu}$ with
$h_{\mu\nu}<0$ for all $\mu ,\nu$.
Thus $\f^{*}h=\sum_{\mu ,\nu =1}^{N}(\f^{*}h_{\mu\nu})
\f^{*}w^{\mu}\otimes \f^{*}w^{\nu}$ 
with $\f^{*}w^{\mu}\in \DP_{1}^+(V)$ by hypothesis and 
$\f^{*}h_{\mu\nu}=h_{\mu\nu}(\f)<0$. Then, $\f^{*}h\in \DP_{2}^-(V)$.

\noindent
$4 \Rightarrow 1$: For every $\u\in\Z^{+}(V)$ we have that 
$(\f^{*}h)(\u,\u)=h(\f^{'}\u,\f^{'}\u)\leq 0$ and hence 
$\f^{'}\u\in\Z(W)$. Besides, for any other $\vec v \in\Z^{+}(V)$, 
$(\f^{*}h)(\u,\vec v)=h(\f^{'}\u,\f^{'}\vec v)\leq 0$ so that every two 
vectors with the same time orientation are mapped to vectors with the 
same time orientation. However, it could happen that $\Z^{+}(V)$ is
actually mapped to $\Z^-(W)$, and $\Z^{-}(V)$ to $\Z^+(W)$. By 
changing the time orientation of $W$, if necessary, the result 
follows.\N

Condition {\em 4} in this theorem can be replaced by

{\em 4'}.\, $\f^{*}h\in \DP_{2}^-(V)$ and $\f^{'}\u\in\Z^+(W)$ for at least 
one $\u\in\Z^+(V)$.

\noindent
Leaving this time-orientation problem aside (in the end, condition {\it 4}
just means that $W$ {\em with one of its time orientations}
is properly causally related with $V$), let us stress that the 
condition {\it 4} (or {\em 4'}) is very easy to check and thereby extremely
valuable in practical problems: first, one only has to work 
with one tensor field $h$, and also as we saw in proposition 
\ref{dp} there are several simple ways to check whether
$\f^{*}h\in\DP_{2}^-(V)$ or not.

The combination of theorem \ref{theo:bb2} and condition {\it 4} in 
theorem \ref{pcr} provides a classification of the proper causal 
relations according to the number of independent null vectors which 
are mapped by $\f$ to null vectors at each point. The key result here 
is that 
\begin{prop}
Let $V\prec_{\f} W$ and $\u\in\Z^{+}_{x},\, x\in V$. Then 
$\f^{'}\u$ is null at $\f(x)\in W$ if and only if $\u$ is a null 
eigenvector of $\f^{*}h|_{x}$.
\label{CONE}
\end{prop}
\P See \cite{GPS}.
Recall that $\u$ is called an ``eigenvector''
of a 2-covariant tensor $T$ if $T(\,\, ,\u )=\lambda g (\,\, ,\u )$ and 
$\lambda$ is then the corresponding eigenvalue.\N

The vectors which remain null under the causal relation $\f$ are 
called its {\em canonical null directions}, and there are at most $N$ 
of them linearly independent. Hence, using theorem \ref{theo:bb2} 
one can see that there essentially are $N$ different types of proper 
causal relations, and that the conformal relations are included as the 
particular case in which {\em all} null directions are canonical 
\cite{GPS}.

Clearly $V\prec V$ for all $V$ (by just taking the identity 
mapping). Moreover
\begin{prop} $V\prec W$ and $W\prec U \Longrightarrow  V\prec U$.
\label{ORDER}
\end{prop}
\P  Consider any $\u\in \Z^{+}(V)$. Since there are $\f ,\psi$ such that
$V\prec_{\f} W$ and $W\prec_{\psi} U$, then $\f^{'}\u\in \Z^{+}(W)$ and
$\psi^{'}[\f^{'}\u]\in\Z^{+}(U)$ so that 
$(\psi\circ\f)^{'}\u\in\Z^{+}(U)$ from where $V\prec U$.\N

It follows that the relation $\prec$ is a preorder for the class of 
all Lorentzian manifolds. This is not a partial order as $V\prec W$ and
$W\prec V$ does not imply that $V=W$ and, actually, it does not even 
imply that $V$ is conformally related to $W$. The point here is 
that $V\prec_{\f} W$ and $W\prec_{\psi} V$ can perfectly happen with 
$\psi \neq \f^{-1}$. In the case that $V\prec_{\f} W$ and 
$W\prec_{\f^{-1}} V$ then necessarily $\f$ is a conformal relation 
and $\f^{*}h=e^{2f}g$, but we are dealing with more general and basic 
causal equivalences.
\begin{defi}
Two Lorentzian manifolds $V$ and $W$ are called causally isomorphic, 
denoted by $V\sim W$, if $V\prec W$ and $W\prec V$.
\label{EQUIV}
\end{defi}
If $V\sim W$ then the causal structures in both manifolds are somehow 
the same. Clearly, $\sim$ is an equivalence relation, and now
one can obtain a partial order $\preceq$ for the corresponding classes of
equivalence coset$(V)\equiv \{U\, : \, \, V\sim U\}$, by setting
$$
\mbox{\rm coset}(V)\preceq \mbox{\rm coset}(W) \Longleftrightarrow 
V\prec W \, .
$$
This partial order provides chains of (classes of equivalence of)
Lorentzian manifolds which keep ``improving'' the causal properties of 
the spacetimes. To see this, we need the following (see \cite{HE,S0} 
for definitions)
\begin{prop}
Let $V\prec W$. Then, if $V$ violates any of the following
\begin{enumerate}
\item the chronology condition,
\item the causality condition,
\item the future-distinguishing condition (or the past one),
\item the strong causality condition,
\item the stable causality condition,
\item the global hyperbolicity condition,
\end{enumerate}
so does $W$.
\end{prop}
\P For {\it 1} to {\it 4}, let $\g$ be a causal curve responsible for 
the given violation of causality (that is, a closed timelike curve 
for {\it 1}, or a curve cutting any neighbourhood of a point in a 
disconnected set for {\it 4}, and so on). Then, $\f(\g)$ has the 
corresponding property in $W$. To prove {\it 5}, if there were a 
function $f$ in $W$ such that $-df\in \DP_{1}^+(W)$, from theorem 
\ref{pcr} point {\it 3} it would follow that 
$d(\f^{*}f)=\f^{*}df\in\DP_{1}^-(V)$ so that $\f^*f$ would also be a 
time function. Finally, {\it 6} follows from corollary 3.1 in 
\cite{GPS}.\N

With this result at hand, we can build the afore-mentioned chains of 
spacetimes, such as
$$
\mbox{\rm coset}(V)\preceq \dots \preceq\mbox{\rm coset}(W)\preceq \dots
\preceq\mbox{\rm coset}(U)\preceq \dots \preceq\mbox{\rm coset}(Z)
$$
where the spacetimes satisfying stronger causality properties are to 
the left, while those violating causality properties appear more and 
more to the right. This is natural because the light cones ``open 
up'' under a causal mapping. The actual properties depend on the particular 
chain and its length, but an optimal one would start with a $V$ which 
is globally hyperbolic, and then it could pass through a $W$ which is 
just causally stable, then $U$ could be causal, say, and the last 
step $Z$ could be a totally vicious spacetime \cite{S0}.

Perhaps the above results can be used to give a first fundamental 
characterization of {\it asymptotically equivalent spacetimes}, at a 
level prior to the existence of the metric, which might then be 
included in a subsequent step. This could be accomplished by means of 
the following tentative definitions, which may need some refinement. 
\begin{defi}
An open set $\z\subset V$ is called a \underline{neighbourhood of}
\begin{enumerate}
\item \underline{the causal boundary} of $V$ if $\z\cap \g\neq \emptyset$ for all 
endless causal curves $\g$;
\item \underline{a singularity set} ${\cal S}$ if $\z\cap \g\neq \emptyset$
for all endless causal curves $\g$ which are incomplete towards 
${\cal S}$;
\item \underline{the causal infinity} if $\z\cap \g\neq \emptyset$
for all complete causal curves $\g$.
\end{enumerate}
\end{defi}
(See \cite{S0} for definition of boundaries, singularity sets, 
etcetera). 
\begin{defi}
$W$ is said to be causally asymptotically $V$ if any two neighbourhoods 
of the causal infinity $\z\subset V$ and $\tilde\z\subset W$ contain 
corresponding neighbourhoods $\z'\subset\z$ and 
$\tilde\z'\subset\tilde\z$ of the causal infinity such that 
$\z'\sim\tilde\z'$.
\end{defi}
Similar definitions can be given for $W$ having causally the 
singularity structure of $V$, or the causal boundary of $V$, 
replacing in the given definition the neighbourhoods of the causal 
infinity by those of the singularity and of the causal boundary, 
respectively. The usefulness of these investigations is still unclear.

\subsection{Causal transformations and generalized symmetries}
Here the natural question of whether the causal relations can be used 
to define a generalization of the group of conformal motions is 
analyzed. To start with, we need a basic concept.
\begin{defi}
A transformation $\f:V\longrightarrow V$ is called 
{\em causal} if $V\prec_{\f}V$.
The set of causal transformations of $V$ is written as $\C(V)$.
\end{defi}
$\C(V)$ is a subset of the group of transformations of $V$. In fact, 
from the proof of proposition \ref{ORDER} follows that $\C(V)$ is
closed under the composition of diffeomorphisms. As the identity map 
is also clearly in $\C(V)$ its algebraic strucuture is that of a 
submonoid, see e.g. \cite{HHL}, of the group of diffeomorphisms of $V$.
However, $\C(V)$ generically fails to be a subgroup, because (see 
\cite{GPS} for the proof):
\begin{prop}
Every subgroup of causal transformations is a group of conformal 
motions.
\label{GROUP}
\end{prop}
From standard results, see \cite{HHL},
one identifies $\C(V)\cap\C(V)^{-1}$ as the 
group of conformal motions of $V$ and there is 
no other subgroup of $\C(V)$ containing $\C(V)\cap\C(V)^{-1}$.
The transformations in $\C(V)\setminus (\C(V)\cap\C(V)^{-1})$ 
are called {\it proper} causal transformations. 

Take now a one-parameter {\it group} of causal transformations
$\{\f_{t}\}_{t\in \r}$. From proposition \ref{GROUP}
it follows that $\{\f_{t}\}$ must be in fact a group of
conformal motions, and its infinitesimal generator is a conformal 
Killing vector, so that nothing new is found here. Nevertheless,
one can generalize naturally the conformal Killings by building 
one-parameter groups of transformations $\{\f_t\}$ such that only 
part of them are causal transformations. Given that the problem arises 
because both $\f_{t}$ and $\f_{-t}=\f^{-1}_{t}$ belong to the family 
and thus they would both be conformal if they are {\it both} causal,
one readily realizes that the natural generalization is to assume 
that $\{\f_{t}\}_{t\in \r}$ is such that
either $\{\f_{t}\}_{t\in \r^{+}}$ or
$\{\f_{t}\}_{t\in \r^{-}}$ is a subset of $\C(V)$, but 
{\em only one of the two}. Any group $\{\f_{t}\}_{t\in \r}$ with this 
property is called a maximal one-parameter submonoid of
proper causal transformations. Of course, the one-parameter submonoid 
can just be a {\it local} one so that the transformations are defined 
only for some interval $I=(-\epsilon,\epsilon)\in \r$ and only those with
$t\in (0,\epsilon)$ (or $t\in (-\epsilon, 0)$) are proper 
causal transformations.

Let then $\{\f_t\}_{t\in I}$ be a local one-parameter submonoid of
proper causal transformations, and assume that $t\geq 0$ provides the 
subset of proper causal transformations (otherwise, just 
change the sign of $t$). The infinitesimal generator of $\{\f_t\}_{t\in I}$
is defined as the vector field 
$$\xiv=\left.\frac{d\f_{t}}{dt}\right|_{t=0}$$
so that for every covariant tensor field $T$ one has
$$
\left.\frac{d(\f^{*}_{t}T)}{dt}\right|_{t=0}=\lie T 
$$
where $\lie$ denotes the Lie derivative with respect to $\xiv$. 
As $\{\f_t\}_{t\geq 0}$ are proper causal transformations, and using 
point {\em 2} in theorem \ref{pcr}, one gets $\f_{t}^{*}T\in\DP^+_{r}$ for 
$t\geq 0$ and for all tensor fields $T\in \DP^+_{r}$. In particular,
\be
\f_{t}^{*}T\,(\u_{1},\dots ,\u_{r})\geq 0 ,\,\,\, \forall \u_{1},\dots 
,\u_{r}\in \Z^+ , \,\, \forall T\in \DP^+_{r}, \,\, t\geq 0,\label{ft}
\ee
from where we can derive the next result.
\begin{lem}
Let $T\in \DP^+_{r}$ and $\k\in \Z^+$ be such that 
$T(\k,\dots,\k)=0$. If $\f_t\in \C(V)$ for $t\in [0,\epsilon)$, then 
$$
(\lie T)(\k,\dots,\k)\geq 0\, .
$$
\label{lem:ft}
\end{lem}
\P
Under the conditions of the lemma, and due to points {\em 2} and {\em 
3} of proposition \ref{dp}, it is necessary that $\k$ is null, that 
is, $\k\in\partial\Z^+$. From formula (\ref{ft}) one obtains
$\f_{t}^{*}T\, (\k,\dots,\k)\geq 0$ for all $t\in [0,\epsilon)$. But 
$\f_{0}$ is the identity transformation, so
$\f_{0}^{*}T\, (\k,\dots,\k)=T\, (\k,\dots,\k)=0$, from where 
necessarily follows that $\f_{t}^{*}T\, (\k,\dots,\k)$ is a 
non-decreasing function of $t$ at $t=0$, that is to say,
$(d/dt)(\f_{t}^{*}T\, (\k,\dots,\k))|_{t=0}\geq 0$.\N
\begin{coro}
Let $\xiv$ be the infinitesimal generator of a local one-parameter
submonoid of proper causal transformations $\{\f_t\}_{t\in I}$
and choose the sign of $t$ such that
$\{\f_t\}_{t\geq 0}\subset \C(V)$. Then
$$
(\lie g) (\k,\k )\leq 0 , , \hspace{1cm} \forall \k\in\partial\Z
$$
\label{signo}
\end{coro}
\P
Obviosuly $g(\k,\k)=0$ for all null $\k$, and also $g\in\DP^-_{2}$,
so lemma \ref{lem:ft} can be applied to $-g$ and the result 
follows.\N

This result is a generalization of the condition for conformal Killing 
vectors ($\lie g \propto g$) and can be analyzed in a similar manner. 
As a matter of fact, the application of the decomposition theorem \ref{theo:bb2}
to $\f_{t}^{*}g\in \DP^-_{2}$ leads to a much stronger result which 
allows for a complete characterization 
of the vector fields $\xiv$ and their properties.
\begin{theo}
Let $\xiv$ be the infinitesimal generator of a local one-parameter
submonoid of proper causal transformations $\{\f_t\}_{t\in I}$
and choose the sign of $t$ such that
$\{\f_t\}_{t\geq 0}\subset \C(V)$. Then there is a function $\psi$ 
such that
$$
\left[\lie g -2\psi g\right]\in \DP_{2}^-. 
$$  
\label{gsym}
\end{theo}
\P From theorem \ref{theo:bb2} and given that $\f_{t}^{*}g\in \DP^-_{2}$ 
for $t\in [0,\epsilon)$ one has
$$
\f_{t}^{*}g=-\sum_{p=1}^N T_{t}\{\Omega_{p}\}=-\sum_{p=1}^{N-1} 
T_{t}\{\Omega_{p}\}+\Psi_{t}^2\, g
$$
where $T_{t}\{\Omega_{p}\}$ are superenergy tensors of simple 
$p$-forms for all values of $t\in [0,\epsilon)$ and $\Psi_{t}$ are 
functions on $V$ with $\Psi_{0}=1$. Then we have
$\f_{t}^{*}g (\u,\vec v)\leq \Psi_{t}^2\, g(u,v)\leq 0$ for all $\u,\vec 
v\in \Z^+$, or equivalently,
$$
\Psi_{t}^{-2}\f_{t}^{*}g (\u,\vec v)\leq g (\u,\vec v)=
\Psi_{0}^{-2}\f_{0}^{*}g (\u,\vec v)\leq 0
$$
from where a reasoning similar to that in lemma \ref{lem:ft}, by 
taking the derivative with repsect to $t$ at $t=0$, gives
$$
\left[\lie g -2\Phi g\right](\u,\vec v)\leq 0 , \hspace{1cm}
\forall \u,\vec v\in \Z^+
$$
where $\Phi\equiv d\Psi_{t}/dt|_{t=0}$.\N

This set of vector fields generalize the traditional (conformal) symmetries and 
the previous theorem together with theorem \ref{theo:bb2} provides 
first a definition of generalized symmetries, and second its full
classification 
because $\lie g -2\psi g$ itself can be written as a sum of 
superenergy tensors of simple $p$-forms. The number of independent null 
eigenvectors of $\lie g -2\psi g$ (ranging from 0 to $N$) gives the 
desired classification, where $N$ corresponds to the conformal 
Killing vectors. This is under current investigation.
It must be remarked that the above theorem does not provide a 
sufficient condition for a vector field to generate locally a 
one-parameter submonoid of causal transformations.

Several examples of generalized Killing vectors in this sense can be 
presented. One of them is a particular case 
of a previous partial generalization of isometries considered in 
\cite{CHS} and called Kerr-Schild vector fields. They are vector 
fields which satisfy $\lie g \propto \ell \otimes \ell$ and 
$\lie \ell \propto \ell$ where $\ell$
is a null 1-form. Obviously, as $\lie g\in\DP_{2}$ this can give 
rise to a one-parameter submonoid of causal transformations. See 
Example 4 in \cite{GPS} for an explicit case of this.

Another interesting example arises by considering the typical 
Robertson-Walker spacetimes $\rw$, the manifold being
$I\times M_{N-1}(\kappa)$ 
where $I\subseteq \r$ is an open interval of the real line with 
coordinate $x^0$ and 
$M_{N-1}(\kappa)$ is the $(N-1)$-dimensional Riemannian space of constant 
curvature $\kappa$, its canonical positive-definite metric being denoted
here by $g_{\kappa}$. The Lorentzian metric in $\rw$ is the warped product
$$
g=-dx^0\otimes dx^0 + a^2(x^0)\, g_{\kappa}
$$
where $a(x^0)>0$ is a $C^2$ function on $I$. Take the diffeomorphisms 
$\f_{t}:\rw \longrightarrow \rw$ which leave $M_{N-1}(\kappa)$ 
invariant (they are the identity on $M_{N-1}(\kappa)$) and act on $I$ 
as $x^0\rightarrow x^0 +t$. It is immediate that
$$
\f_{t}^{*}g = -dx^0\otimes dx^0 + a^2(x^0+t)\, g_{\kappa}
$$
so that $\f_{t}^{*}g\in \DP^-_{2}(\rw)$ if and only if $a(x^0+t)\leq 
a(x^0)$, and therefore $\f_{t}^{*}g\in \DP^-_{2}(\rw)$ for $t\in 
[0,\epsilon)$ if and only if $a$ is a non-increasing function.
Physically this means that $\{\f_t\}_{t\in I}$ is a one-parameter 
submonoid of proper causal transformations in $\rw$ if and only if the 
Robertson-Walker spacetime is non-expanding. Naturally, the 
non-contracting case, perhaps of more physical importance, can be studied 
analogously by simply changing the sign of $t$.

The infinitesimal generator of this one-parameter group is
$$
\xiv \equiv 
\left.\frac{d\f_{t}}{dt}\right|_{t=0}=\frac{\partial}{\partial x^0}
$$
and the deformation of the metric tensor reads
$$
\lie g = 2a\dot a \,\, g_{\kappa} = \frac{2\dot a}{a}\,\, (g+\xi \otimes 
\xi)
$$
where $\dot a$ is the derivative of $a$ and $\xi=g(\,\, ,\xiv)=dx^0$. 
Observe that,
$$
\lie g =\frac{2\dot a}{a}\, T\{\xi \}+\frac{\dot a}{a}\, g
$$
where $T\{\xi \}$ is the superenergy tensor of $\xi$. Obviosuly, the sign of
$\dot a$ is determinant here for $\lie g -(\dot a/a)\, g$ to be in 
$\DP_{2}$, in accordance with the previous 
reasoning and the theorem \ref{gsym}. In fact, in this explicit 
case, as $g_{\kappa}$ is a positive-definite metric, one can prove 
$$
(\lie g) (\x,\x ) = 2a\dot a \,\, g_{\kappa}(\x,\x ), \,\,\,\, 
\forall \x\in T(\rw)
$$
which has the sign of $\dot a$ for {\em all} vector fields $\x$. 
This same property is shared by the Example 4 of \cite{GPS}. 

All in all, the deformation $\lie g$ produced by 
one-parameter local submonoids of causal transformations
has been shown to be controllable and the generalized symmetries thereby defined
can be attacked using traditional techniques.


{\section*{Acknowledgments}}
\noindent 
I am grateful to the organizers of the meeting for inviting me to such 
an exciting exchange of ideas and information between mathematicians 
and physicists. Some parts of this contribution arise from 
collaborations with Alfonso Garc\ii a-Parrado and G\"oran Bergqvist.
This work is supported by the research project UPV 
172.310-G02/99 of the University of the Basque Country.
\bigskip






\begin{thebibliography}{99}
    
\bibitem{BS} \sc G. Bergqvist and J.M.M.Senovilla, {\em Null cone 
preserving maps, causal tensors and algebraic Rainich theory},
\rm Class. Quantum Grav. {\bf 18}, 5299-5325, (2001).  

\bibitem{Ple} \sc J.F.Pleba\'{n}ski, {\em The algebraic structure of 
the tensor of matter}, \rm Acta Phys. Pol. {\bf 26}, 963-1020 (1964).

\bibitem{HE} \sc S.W.Hawking and G.F.R.Ellis,
\rm {\em The large-scale structure of spacetime},
Cambridge University Press, Cambridge (1973).

\bibitem{S} \sc J.M.M.Senovilla, {\em Super-energy tensors}, \rm Class. 
Quantum Grav. {\bf 17}, 2799-2842 (2000).

\bibitem{S1} \sc J.M.M. Senovilla, {\em Remarks on superenergy 
tensors}, \rm In {\em Gravitation and Relativity in General.
Proc. Spanish Relativity Meeting
in Honour of the 65th Birthday of L Bel} Salamanca 1998,
eds. J.Mart\'{\i}n, E.Ruiz, F.Atrio and A.Molina, pp.175-182,
World Scientific, Singapore (1999). 

\bibitem{Bel} \sc L. Bel, {\em Introduction d'un tenseur du quatri\`eme ordre},
\rm C.R. Acad Sci. Paris {\bf 248}, 1297-1300 (1959).

\bibitem{B} \sc L. Bel, {\em Sur la radiation gravitationnelle},
\rm C.R. Acad Sci. Paris {\bf 247}, 1094-1096 (1958)

\bibitem{Bel2} \sc L. Bel, {\em Les \'etats de radiation et le 
probl\`eme de l'\'energie en relativit\'e g\'en\'erale},
\rm Cahiers de Physique {\bf 16}, 59-81 (1962);
English translation: Gen. Rel. Grav. {\bf 32}, 2047-2078 (2000).

\bibitem{MTW} \sc C. W. Misner, K. S. Thorne and J. A. Wheeler, 
\rm {\em Gravitation}, Freeman, San Francisco (1970). 

\bibitem{R} \sc G.Y.Rainich, \rm  {\em Electrodynamics in general relativity},
Trans. Amer. Math. Soc. {\bf 27}, 106-136 (1925).

\bibitem{MW} \sc C.W.Misner and J.A.Wheeler, {\em Classical physics 
as geometry: Gravitation, electromagnetism, unquantized charge, and 
mass as properties of curved empty space}, \rm Ann.Phys. NY {\bf 2},
525-603 (1957).

\bibitem{PR} \sc R.Penrose and W.Rindler, \rm {\em Spinors and spacetime},
vol.1-2, Cambridge University Press, Cambridge (1986).

\bibitem{L} \sc D.Lovelock, \rm {\em Dimensionally dependent 
identities}, Proc. Camb. Phil. Soc. {\bf 68}, 345-350 (1970).

\bibitem{CF}\sc  B.Coll and J.J.Ferrando, \rm {\em Thermodynamic perfect 
fluid. Its Rainich theory}, J. Math. Phys. {\bf 30}, 2918-2922 (1989).

\bibitem{BerS} \sc G.Bergqvist and J.M.M.Senovilla, {\em On the causal 
propagation of fields}, \rm Class.Quantum Grav. {\bf 16}, L55-L61 (1999).

\bibitem{S0} \sc J. M. M. Senovilla {\em Singularity Theorems and 
their Consequences}, \rm  Gen. Rel. Grav. {\bf 30}, 701-848, (1998).

\bibitem{BoS} \sc M.\'{A}.G.Bonilla and J.M.M.Senovilla {\em Very 
simple proof of the causal propagation of gravity in vacuum}, \rm Phys. 
Rev. Lett. {\bf 78}, 783-786 (1997).

\bibitem{S2} \sc J.M.M. Senovilla, \rm {\em (Super)$^n$-Energy for
Arbitrary Fields and its Interchange: Conserved Quantities},
Mod. Phys. Lett. A {\bf 15}, 159-165 (2000).

\bibitem{MS} \sc M. Mars and J.M.M. Senovilla, \rm {\em Geometry of 
general hypersurfaces in spacetime: junction conditions}, Class. 
Quantum Grav. {\bf 10}, 1865-1897 (1993).

\bibitem{Fri} \sc F.G.Friedlander, \rm {\em The wave equation on a curved
spacetime}, Cambridge University Press, Cambridge (1975).

\bibitem{Li} \sc A. Lichnerowicz, \rm {\em Ondes et radiations
\'electromagn\'etiques et gravitationnelles en relativit\'e g\'en\'erale},
Ann. di Mat. Pura ed Appl. {\bf 50}, 1-95 (1960).

\bibitem{Had} \sc J. Hadamard, \rm {\em Le\c{c}ons sur la propagation
des ondes et les \'equations de l'hydrodynamique}, Hermann, Paris, (1903).

\bibitem{Ge} \sc G. Gemelli, \rm {\em Gravitational waves and
discontinuous motions}, Gen. Rel. Grav. {\bf 29}
161-178 (1997).

\bibitem{GPS} \sc A.Garc\ii a-Parrado and J.M.M. Senovilla, \rm {\em 
Causal relations and their applications}, contribution to this volume.

\bibitem{HHL} \sc J. Hilgert, K. H. Hofmann and J. D. Lawson 
\rm {\em Lie groups, Convex Cones and Semigroups.} Oxford Sciencie 
Publications, (1989). 

\bibitem{CHS} \sc B. Coll, S. R. Hildebrandt and J. M. M. 
Senovilla. \rm {\it Kerr Schild Symmetries.} Gen. Rel. Grav. {\bf 
33}, 649-670, (2001).

\end{thebibliography}
\end{document}